\documentclass[aps,prb,twoside,twocolumn,showpacs,groupedaddress,floatfix]{revtex4-1}  
\usepackage{graphicx}  
\usepackage{dcolumn}   
\usepackage{amsmath}
\usepackage{amssymb}   
\usepackage{color}
\usepackage[normalem]{ulem}

\usepackage{times}

\renewcommand{\O}[0]{{\text{O}}}

\begin{document}



\title{Magnetic fluctuations from oxygen deficiency centers on the SiO$_2$ surface}

\author{Nicole Adelstein}
\email{adelstein1@llnl.gov}
\author{Donghwa Lee}
\author{Jonathan L. DuBois}
\author{Vincenzo Lordi}
\email{lordi2@llnl.gov}
\affiliation{Lawrence Livermore National Laboratory, Livermore, CA, USA, 94550}

\date{\today}

\begin{abstract}
The magnetic stability of oxygen deficiency centers on the surface of $\alpha$-quartz is investigated with first-principles calculations to understand their role in contributing to magnetic flux noise in superconducting qubits (SQs) and superconducting quantum interference devices (SQUIDs) fabricated on amorphous silica substrates.  Magnetic defects on the substrate are likely responsible for some of the 1/f noise that plagues these systems.  Dangling-bonds associated with three-coordinated Si atoms allow electron density transfer between spin up and down channels, resulting in low energy magnetic states.  Such under-coordinated Si defects are common in both stoichiometric and oxygen deficient silica and quartz and are a probable source of magnetic flux fluctuations in SQs and SQUIDs.
\end{abstract}

\maketitle


\section{\label{sec:level0} Introduction}

The atomic source of magnetic flux fluctuations in quantum interference devices (SQUIDs) and superconducting qubits (SQs)\cite{wellstood87, bialczak07, bluhm09, bylander11, chang13, agarwal13, lee13} is still not well understood\cite{anton13}.  Engineering the material components of SQs and SQUIDS to remove magnetic defects could help reduce the 1/f noise from magnetic flux fluctuations and increase quantum states' coherence times, which would enable the development of scalable solid-state quantum computers \cite{oliver13}.  Experimental efforts focused on the dependence of the noise spectral density on device geometry \cite{bialczak07, bylander11} suggest that the source of this 1/f noise lies in the substrate surface or interfaces between the superconducting material and the substrate or the superconductor's oxide surface.  

The goal of this first-principles study is to identify and characterize defects with low energy magnetic states (LEMS) on the surface of the silica substrate of SQs and SQUIDS.  Previously, LEMS have been associated with adsorbates on the sapphire substrate using first-principles simulations and alternatives to displace these defects were proposed\cite{lee13}.  In this work, we expand upon the sapphire study by focusing on defects on the surface of silica, a common and inexpensive substrate material for SQUIDs and SQs \cite{martinisPRL05, schickfus77, oconnell08}, and in particular on defects associated with the oxygen vacancy ($V_{\O}$) or oxygen deficiency center (ODC). 

Previous SQUID and SQ research has proposed that magnetic flux fluctuations are caused by unpaired electrons hopping between defect sites \cite{kochPRL07} or flipping spin\cite{desousa07}.  Both these studies point to paramagnetic $V_{\O}$ defects associated with Si dangling-bonds as a possible of source of hopping or flipping electrons, but do not explicitly compute electronic or magnetic properties of these defects in the bulk or on the surface of silica.  

Paramagnetic defects in the bulk and on the surface of silica have been extensively studied \cite{trisi, skujaPSS05} because they degrade the performance of silica in optics and integrated circuits.  Previous calculations \cite{pacchioni98} have assigned the undesired optical transitions in SiO$_2$ to specific defect geometries, including the Si dangling-bond from an oxygen deficiency center, which is a more generalized term for the oxygen vacancy in both stoichiometric and non-stoichiometric SiO$_2$.   While the electronic structure and geometries of ODCs have been determined with first-principles calculations, little is known about their magnetic properties on the surface of silica.  

Here, we present a computational investigation into the magnetic properties of SiO$_2$ surfaces and both paramagnetic and diamagnetic defects on the surface. Using density functional theory (DFT) simulations of $\alpha$-quartz as a model for amorphous silica, we have identified intrinsic defects with local magnetic moments on SiO$_2$ surfaces that are associated with Si dangling-bonds. The geometries of these common defects are the Si--Si dimer and puckered configuration (PC), which are described in the next section.  We discuss how a local magnetic moment from a Si dangling-bond can cause magnetic noise Since these Si dangling-bonds are associated with ubiquitous E$^\prime$ paramagnetic defects in silica \cite{andersonPRL11, andersonAPL12, skujaPSS05, sushko05}, they are a probable source of 1/f noise in SQs and SQUIDs. 

\section{\label{sec:level1} Simulation Details}

\subsection{\label{sec:1A} Computational Methods}
We employ DFT \cite{hohenberg} with the local spin density approximation (LSDA) to study the thermodynamic and magnetic properties of surface defects.
The Vienna \textit{Ab-Initio} Simulation Package (VASP) with the projector augmented-wave method is used\cite{ perdew81, kresse93, kresse96, blochl}.  The calculations are converged with respect to plane wave cutoff and $k$-point sampling using the $\alpha$-quartz unit cell, which contains 2 formula units (f.u.).  A 6$\times$6$\times$6 Monkhorst-Pack \cite{monkhorst} $k$-point grid centered at $\Gamma$ gives energy convergence to $2.2\times10^{-5}$~eV for the unit cell.
 A plane wave cutoff of 700~eV gives convergence to 1~meV.  The forces on the ions are relaxed to 0.01~eV/\AA.
 
  Positively charged states are created by removing electrons and compensating with a homogeneous negative background charge by setting the $G=0$ component of the potential to zero. Magnetic excited states are created by constraining the difference between the spin up and spin down populations.
 
 \subsection{\label{sec:1B} Defect and Surface Structures}
 
$\alpha$-quartz is hexagonal with the P3$_2$31 space group (\#154) and we simulate the left-handed chirality.  Our calculations give a lattice parameter of 4.884~\AA, $c/a$ ratio of 1.1023, and internal coordinates of $\left(0.4658, 0.0, 0.0\right)$ and $\left(0.4120, 0.2740, 0.1141\right)$ for Si and O, respectively, which match the experimental and previous computational values well \cite{gnani, demuth99, levien80}.

\begin{figure}
\includegraphics[trim={40 20 0 130}, scale=0.35]{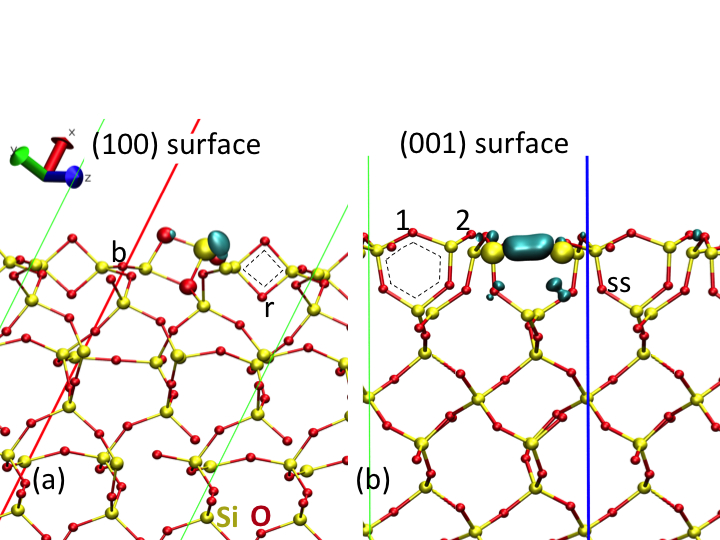}
\caption{\label{fig:empty}(a) The puckered configuration on the (100) surface of $\alpha$-quartz with the $V_{\O}$ created from the 2Si ring, and (b) the Si--Si dimer created from a $V_{\O}$ on the (001) surface are highlighted with larger atomic spheres.  The thick red line is the [100] axis and the thick blue line is the [001] axis, while the thin green lines demarcate the edges of the supercells that were used.  The blue isosurfaces show the localized empty states from the Si dangling bond on the (100) surface in (a) and from the Si--Si bond on the (001) surface in (b). The labels (b, r, 1,2, and ss) refer to different positions of the $V_{\O}$ that were tested.  Ring structures are highlighted with dashed lines. 
}
\end{figure}

In bulk quartz, Si has a coordination number of four, but an oxygen deficiency center can create three-coordinated Si atoms (III-Si) in either the Si--Si dimer or puckered configuration\cite{sushko05}, which are shown in Fig. \ref{fig:empty}. Both the PC and the charged dimer have LEMS with localized magnetic moments due to the unpaired electron in the dangling bond of the PC or the between the two Si atoms of the dimer; both defects are common bulk and surface defects of silica. 

 The dimer configuration is a Si--Si bond, labeled the ODC(I) in the literature.  While there are many possible intrinsic defects on the surface of $\alpha$-quartz and silica, the ODC(I) has nearly identical optical characteristics in both silica and $\alpha$-quartz \cite{skujaPSS05}, making quartz a reasonable model for the ODC(I)\ on the silica surface. The neutral oxygen vacancy (NOV) is often associated with the dimer configuration, since the removal of a neutral oxygen atom leaves behind two electrons from the Si--O bonds that can form a Si--Si bond.  

The puckered configuration involves a Si atom with only three oxygen bonds and has a dangling $sp^3$ orbital.   The PC has been labeled the ODC(II) and has been modeled extensively with first-principles calculations \cite{andersonPRL11, andersonAPL12, uchino06, stirling02, pacchioni98ig, vanginhoven03}. The PC can be considered a relaxation of the dimer, where one of the Si atoms in the dimer bends towards an oxygen atom in the bulk to create a new Si--O bond, leaving the other Si with a dangling bond.  

 The dimer and PC can be simulated with the (001) and (100) $\alpha$-quartz surfaces, respectively. The chains of two-Si rings on the (100) surface, as shown in Fig. \ref{fig:empty} (a), are representative of the (101) and (10$\bar{1}$) surfaces, while the (001) surface has a unique honeycomb pattern of six-Si membered rings. Realistic surface models, which are periodic in 2 dimensions with a vacuum region in the third, require a slab thick enough to contain a bulk-like center.  We converge our calculations with respect to the slab thickness and the vacuum size in order to confirm that the defects on the surfaces are screened by both the bulk-like region in the center of the slab and the vacuum. Finite size convergence studies of defect formation energies and magnetic stabilities of one of the defects showed that a simulation cell with a surface area of $\sim$100~\AA$^2$/side and a vacuum spacing of 15~\AA\  is adequate to model the dilute limit.  The dimensions of the supercells used in our calculations are summarized in Table \ref{tab:surfaceE}. The cell shape is allowed to change when relaxing the slabs, but the volume is held fixed.

\section{\label{sec:level3}Results}
 
We identify LEMS by calculating the magnetic stability of charged defects on the (001) and (100), which are representative of a variety of local environments that would be found on a SQ or SQUID silica substrate. We find that an unpaired electron on a III-Si can be stabilized compared to paired electrons due to the lack of screening from the electron with opposite spin and the Coulomb repulsion between the electrons.  Thus, even for diamagnetic dimer and PC defects, which can be simulated with a $V_{\O}^{2+}$, a low energy state with unpaired electrons is accessible. 

 The defect formation energies for a subset of 18 representative oxygen vacancies in different charge states and geometries are presented to indicate which charge states are more thermodynamically favorable.  These results shed light on the relationship between the thermodynamic stability of the defect's geometry and the magnetic stability.   For completeness, we first report our calculation of the low energy reconstructed $\alpha$-quartz surfaces and their magnetic stability. 

\subsection{\label{sec:levelA}Surface energies}
The (001) and (100) surfaces of $\alpha$-quartz represent a variety of local environments which are found on experimental silica surfaces \cite{skujaPSS05}.  A number of previous computational studies have investigated the surface energy of different $\alpha$-quartz surfaces, but there is not a clear consensus on which is the lowest energy surface \cite{deLeeuw, murashovJPCB05}.
Table \ref{tab:surfaceE} summarizes our computed surface energies.
We find that the reconstructed (001) surface has the lowest energy, similar to De Leeuw \emph{et al} \cite{deLeeuw}.  

\begin{table}
\caption{\label{tab:surfaceE} Calculated surfaces energies and slab lattice parameters for five surfaces of $\alpha$-quartz.
}
\begin{ruledtabular}
\begin{tabular}{lccccc} 
& Energy & $a$ & $b$ & $c$ \\
Surface & [meV/\AA$^2$] & [\AA] & [\AA]  & [\AA] \\
(001) &  40.8 &  (4.88,0.0,0.0) &  (0.0,8.46,0.0) & (0.0,0.0,37.69) \\
(101) & 68.7 & (4.88,0.0,5.38) & (-2.44,4.23,0.0) & (-34.19,0.0,37.69) \\
(100) & 116 & (48.84,0.0,0.0) & (-7.33,12.69,0.0) & (0.0,0.0,10.77) \\
(100)$^*$ & 158 &  &  & \\
$(10\bar{1})^*$ & 129.8 & (4.88,0.0,-5.38) &  (-2.44,4.23,0.0) &  (29.31,0.0,32.3)\\
\end{tabular}
\end{ruledtabular}
$^*$Starred surfaces have under- or over- coordinated atoms.
\end{table}

In our simulations, the surfaces are cut along specific lattice planes, ensuring stoichiometric slabs, and then the broken dangling bonds are allowed to relax.   
Our converged (001) slab has 8 Si layers (8 f.u.); the (101) slab has 9 Si layers (27 f.u.); the (10$\bar{1}$) slab has 7 Si layers (21 f.u.); the (100) surface has 7 Si layers (20 f.u.). 
 
The surfaces relax to give the bulk coordination number for both Si and O in the (001), (101) and (100) slabs, but not for the starred (100)$^*$ surface and $(10\bar{1})^*$ surfaces. On these two surfaces, which are higher in energy, some of the cleaved bonds are not able to fully reconstruct before the forces on the atoms coverage 0.01 eV/\AA, leaving (100)$^*$ with dangling oxygen bonds and the $(10\bar{1})^*$ surface with oxygen coordinated by 3 Si atoms (III-O). 
All surfaces other than the (001) have chains of 2 Si membered rings making a diamond like shape, with one of the two oxygen atoms pointing out of the surface, as shown in Fig. \ref{fig:empty} (a) with a dashed diamond and the lower oxygen labeled ``r'', for ring.  We call these ``2Si rings", which are highly strained and thus susceptible to rupture by H$_2$O molecules or other adsorbates.

\subsection{\label{sec:level4}Defect formation energies}

Oxygen vacancies are created from all possible oxygen environments on the (001) and (100) surfaces and the formation energies are calculated in order to estimate the probability of finding these defects.  The (001) surface in Fig. \ref{fig:empty}(b) shows the two distinct surface oxygen atoms (labeled 1 and 2)  and the subsurface atom (labeled ss) that are removed to create $V_{\O}$. While indistinguishable in the honeycomb pattern on the surface, in the subsurface one oxygen is part of a three-membered Si ring, indicated with a dashed line in Fig. \ref{fig:empty} (b), while the other is not (labeled 1 and 2 respectively).   Four sites are tested on the (100) surface, including both the oxygen in the 2Si ring and the two distinct oxygen atoms that bridge the rings in the chain.  

We calculate the magnetic stability of the three (001) $V_{\O}$ and three of the lower energy vacancies on the (100) surface, one $V_{\O}$ from a bridging oxygen and the $V_{\O}$ from the ring oxygen that points into the surface, both the metastable and stable geometries.  Seven sites are tested on the (100)$^*$ surface, but all the formation energies for these defects are negative, indicating that the surface is not stable and thus we do not consider them to be oxygen vacancies.

\begin{figure}
\includegraphics[trim={0 30 0 0}, scale=0.34]{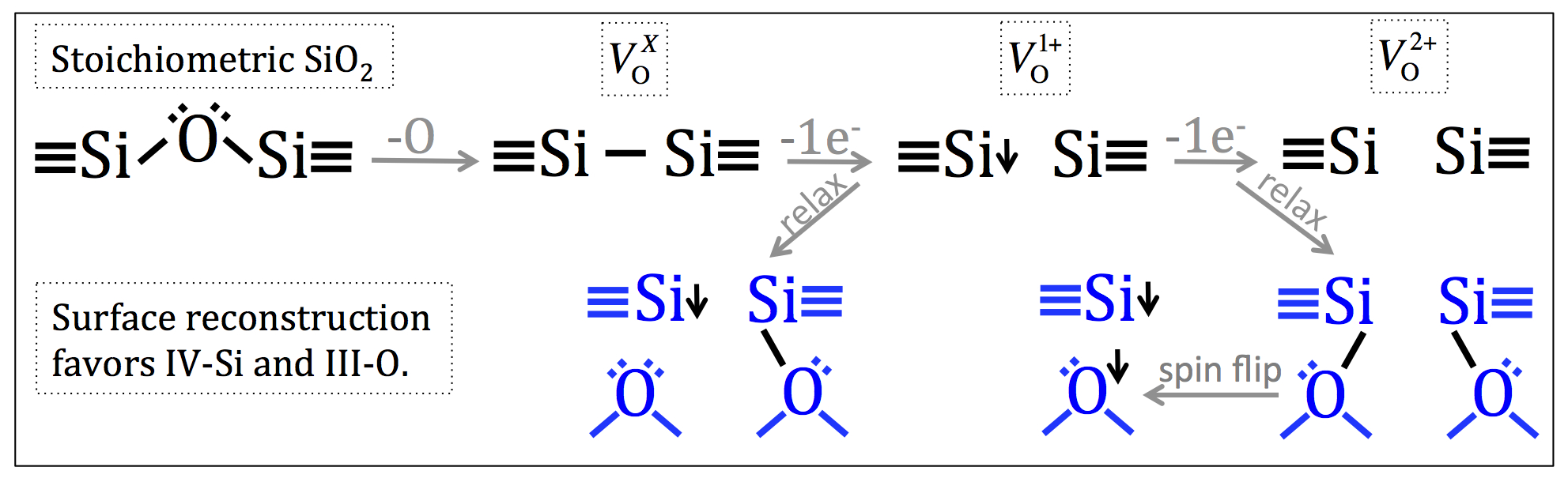}
\caption{\label{lewisdot} {A Lewis dot structure schematic shows the possible bonding configurations for the neutral versus positively charged oxygen deficiency centers. The three stacked lines indicate the additional Si--O bonds of Si's tetrahedral coordination. The top row illustrates the oxygen vacancy where no bond reformation occurs, while the bottom row shows a neighboring oxygen atom bonding with Si to create the puckered configuration.}
}
\end{figure}

We are able to model the ODC dimer configuration on the (001) surface and the puckered configuration (PC) with the charged bridge $V_{\O}$. The neutral oxygen vacancy (NOV) creates a Si--Si dimer in all environments on the (001) and (100) surfaces, as diagrammed in Fig. \ref{lewisdot}, and is a color center called the ODC(I) in the spectroscopy community \cite{skujaPSS05, andersonPRL11}. Both the III-Si and II-Si have been identified as the ODC(II) and have at least one Si dangling bond, which can be modeled using the (100) bridge $V_{\O}$ as seen in Fig. \ref{fig:empty} (a).   Both ODC(I) and ODC(II) are potential precursors of paramagnetic E$^\prime$ centers \cite{andersonPRL11, boeroPRL03}.

\begin{table*}
\caption{\label{tab:Eform}Formation energies (in~eV) for 18 oxygen vacancy defects show similarities between certain bonding environments.  
Defect geometries are Si--Si dimer, unless otherwise indicated in parentheses; the defects in the last column all have the same 2III-O geometry.
Entries in \textbf{bold} have LEMS.
}
\begin{ruledtabular}
\begin{tabular}{lccccccc}
   		&  			&	(001) $V_{\O}$		&				& \vline &						&    (100) $V_{\O}$		&   \\
Charge	& 	site 1 		& site 2 	  		& subsurface		& \vline &    bridge 		 		  	& metastable ring		& ring 2III-O \\ \hline
0 		& 5.1378 			& 5.2306    		& 5.5779			& \vline &    5.2099			  	  	& 5.3224 				& 3.4193 \\
1+ 		& \textbf{5.3593}  	& \textbf{5.3537}    	& \textbf{5.5137} 			& \vline &    \textbf{3.6295} (PC)   	 	& \textbf{3.1666} (PC) 	& \textbf{0.6417} \\
2+ 		& \textbf{5.9628}  	& \textbf{5.9717}	& 2.7874 (2III-O)	& \vline &    \textbf{3.6239} (PC)	& 1.2963 (2III-O/V-Si)	& -1.1842 \\
\end{tabular}
\end{ruledtabular}
\end{table*}

Table \ref{tab:Eform} gives the formation energies, $E_{\text{form}}$, under electron withdrawing conditions when the Fermi level ($E_{\text{F}}$) is at the valence band maximum (VBM), as given in Eq. \ref{eq:eform} below:
\begin{equation}
\label{eq:eform}
E_{\text{form}} = E_{\text{defect}}-E_{\text{slab}}+\frac{1}{2} \mu_{O_2} + q(E_{\text{VBM}}),
\end{equation}
where $E_{\text{defect}}$ is the total energy of the slab containing the defect,  $E_{\text{slab}}$ is the total energy of the stoichiometric reconstructed slab,  
$\mu_{O_2}$ is the chemical potential of molecular oxygen, and $q$ is the charge of the defect.
In our calculations, $E_{\text{VBM}}$ for the (100) slab is $-1.8181$~eV and for the (001) slab is $-2.6358$~eV; 
 $\mu_{\O_2}$ is calculated to be $-10.4857$~eV with LSDA.

Almost all of the defects tested have positive formation energies in oxidizing conditions, \textit{i.e.} when the chemical potential of oxygen is that of molecular O$_2$ in its standard state and $\mu_{O}=\frac{1}{2}\mu_{O_2}$. In complex oxides with ionic-like bonds, the oxygen vacancy is often more stable in the 2+ charge state, but SiO$_2$ has very covalent-like bonds and we calculate that the neutral oxygen vacancy is preferred when the Fermi level is in the middle of the gap.  
Only the stable ring vacancy in the 2+ charge state, where all Si are tetrahedrally coordinated to oxygen, has a negative formation energy (last column of Table \ref{tab:Eform}).

On both surfaces, the Si--Si dimer in the neutral, 1+, and 2+ charge states has a formation energy between 5 and 6~eV ($E_{\text{form}}$ without geometries specified in parentheses are dimers in Table \ref{tab:Eform}). Removing a neutral oxygen breaks two Si--O bonds and the two remaining valence electrons from the Si atoms can form a Si--Si bond, as indicated in Fig. \ref{lewisdot}.  All neutral $V_{\O}$ accommodate Si--Si bonds, except for the stable ring defect in the last column of Table \ref{tab:Eform}, which is relaxed from the 2III-O geometry, as described further below. 

In general, as electrons are removed from the Si--Si bond, the formation energy increases or the bonds rearrange to lower the energy by creating four-coordinated, tetrahedral Si atoms.  On the (001) surface, where no bond rearrangement occurs, empty defect states are created when the electrons are removed from the dimer, leaving III-Si atoms, and the formation energy increases (see the top row of Fig. \ref{lewisdot} and columns 1 and 2 in Table \ref{tab:Eform}).  The hexagonal ring structure of the (001) surface is very stable and thus does not undergo bond rearrangement to lower the defect energy of the III-Si by creating new Si--O bonds using the electrons from lone pairs of neighboring oxygen atoms.

The puckered configuration forms only one new Si--O bond and leaves a dangling III-Si.  This geometry has a formation energy around 3-4~eV  and  occurs for the charged bridge $V_{\O}$ and the metastable ring $V_{\O}^{1+}$ on the (100) surface (a Lewis dot schematic for the PC is given in the lower left of Fig. \ref{lewisdot}).  Upon further atomic rearrangement of atoms, the dangling III-Si atom can bond with a different O neighbor and create the 2III-O configuration with formation energies less than $\sim$3~eV (a Lewis dot schematic for this configuration is given in the lower right corner of Fig. \ref{lewisdot}).  The bottom Si of the 2Si ring and the III-Si in the subsurface of the (001) slab have many oxygen neighbors, so they can find a O neighbor to form a Si--O bond. 

For the ring $V_{\O}^{2+}$  on the (100) surface, the 2III-O/V-Si configuration is a metastable strained geometry with maximum forces of 0.015~eV/\AA , which then relaxes to the stable ring defect with 2III-O atoms.   The last entry of the last column of Table \ref{tab:Eform} shows that removing all III-Si atoms results in negative $E_{\text{form}}=-1.2$~eV, which is significantly lower than the metastable $E_{\text{form}}$ of 1.3~eV.  The 2III-O configuration is tested in the 1+ and neutral charge states by adding electrons back to the $V_{\O}^{2+}$ supercell; allowing the cell to relax does not alter the 2III-O configuration.  All atoms have complete octets in the 2III-O configuration in the 2+ charge state, so adding electrons back to the cell significantly increases $E_{\text{form}}$, though the 2III-O configuration is always more stable than the metastable ring configuration with under and over coordinated Si atoms (compare the last two columns of Table \ref{tab:Eform}). 

In summary, the Si--Si bond is destabilized as electrons are removed from the dimer.  Positively charged defects can occur due to charge trapping or can be charge compensated by negatively charged defects.  Forming new Si--O bonds from the lone pairs of neighboring oxygen atoms can stabilize the defects, resulting in complete octets for the Si bonded to the III-O. 

We note that, all the defects that we have studied may be frozen into place in a glass, even though they have high formation energies in the crystal. Also, while the formation energies are high in oxygen rich conditions, Si rich conditions give an oxygen chemical potential about 4~eV lower in energy, as calculated by Boero et al.\cite{boeroPRL03}, which would make the formation energies of most of these defects much more accessible and in fact they are recognized as the ODC(I) and E$^\prime$ defects in spectroscopic experiments. 

\subsection{\label{sec:level5}Magnetic stability}

The reconstructed surfaces are magnetically stable (have no LEMS).  For example, the smallest energy difference between the non-magnetic ground state and one with a magnetic moment of 1~$\mu_{\text{B}}$ is 1.9~eV for the (100)$^*$ surface.    

Each of the $V_{\O}$ formation energies in Table \ref{tab:Eform} is given for the lowest energy magnetic state.  For the $V_{\O}^{1+}$, the lowest energy state has a magnetic moment of 1~$\mu_{\text{B}}$ on the (100) and (001) surfaces.  For the neutral and 2+ defects, the lowest energy state is non-magnetic, except the 2+ bridge defect on the (100) surface, which is unique in having a magnetic ground state of $\sim$1~$\mu_{\text{B}}$.  From the set of 18 oxygen vacancies that are examined, all $V_{\O}^{1+}$ have a ground state magnetic moment and three $V_{\O}^{2+}$ have LEMS.  These are the 2+ charged $V_{\O}$ from sites 1 and 2 on the (001) surface and the $V_{\O}^{2+}$ bridge vacancy on the (100) surface (shown in bold font in Table \ref{tab:Eform}).  

\begin{figure}
\includegraphics[trim={150 80 90 60}, scale=0.35]{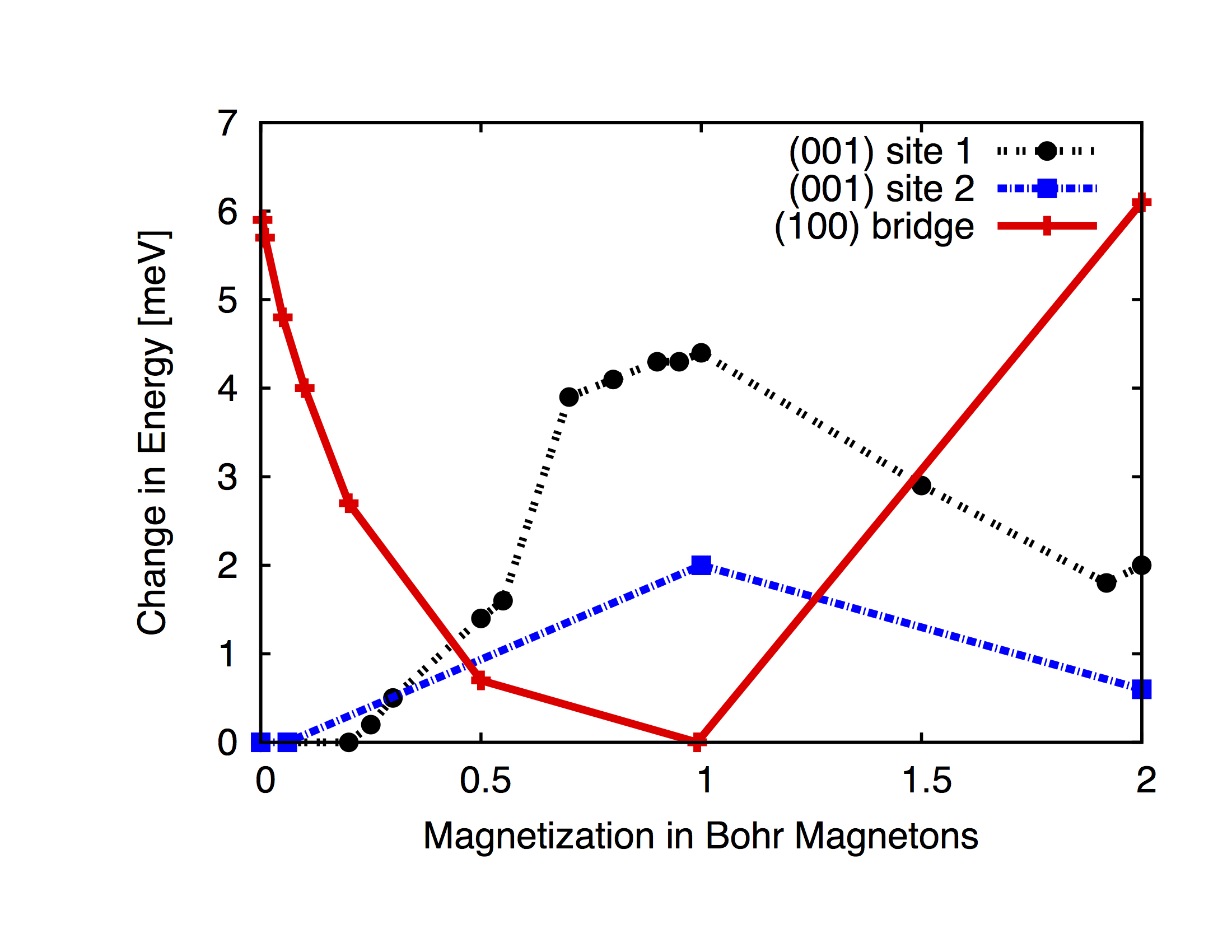}
\caption{\label{fig:mag} The difference in energy from the magnetic ground state as a function of magnetization is shown for three of the five LEMS defects that have been identified. ``Site 1'' and ``Site 2'' refer to 2+ charged dimer defects on the (001) surface and ``bridge'' refers to the $V_{\O}^{2+}$ from the (100) surface. (See Table \ref{tab:Eform}.)
}
\end{figure}

In Fig. \ref{fig:mag}, the energy versus the magnetic moment is shown for the $V_{\O}^{2+}$ at both sites on the (001) surface and the (100) bridge $V_{\O}^{2+}$.  For the (001) $V_{\O}^{1+}$ defect, which is not shown in Fig. \ref{fig:mag}, the non-magnetic state is 0.0384~eV higher in energy than the ground state with magnetic moment of 1~$\mu_B$.  

The approximation of collinear spins in the LSDA  allows for a number of ambiguous magnetizations to be tested, such as a paramagnetic system with a magnetization of 0.5 $\mu_B$.  These magnetizations are due to the partial occupancy of spin up versus spin down eigenstates, usually near the top of valence band.  We use these partial magnetizations to transition between the ground and excited states without becoming trapped in other metastable minima.  

Note that finding the non-magnetic (singlet) ground state is non-trivial on the (001) surface, since the ground state is spin asymmetric and the DFT code does not automatically break the symmetry of the spin channels. The metastable spin symmetric state is $\sim$0.01~eV higher in energy than the nominally non-magnetic spin asymmetric ground state.   The ground state can be found by manually setting the initial occupations of eigenstates to mimic the asymmetric configuration before relaxation \emph{and} enforcing a zero magnetization.  

\section{\label{sec:level6}Discussion}

We find that LEMS are associated with under-coordinated III-Si atoms, which can have local magnetic moments due to the stability of the unpaired electron in the dangling bond on the III-Si.  These III-Si defects are found on both surfaces tested: from the Si--Si dimer in the 1+ and 2+ charge states on the (001) surface and from the PC's dangling bond that points out of the (100) surface. Recall that we broadly define LEMS to be either defects with a magnetic ground state or a low energy magnetic state, and that the oxygen vacancies on (001) and (100) surfaces are representative models for the kinds of defects that exist in silica. 

LEMS associated with the III-Si have defect states in the band gap with energies very close to $E_{\text{F}}$.  In contrast, stable non-magnetic defects have stable spin symmetric electronic structures which require significant energy to create unpaired electrons by breaking spin symmetry.  It is illuminating to first examine the electronic structure of these stable non-magnetic defects. 

\subsection{\label{sec:level7}Non-magnetic defects}

\begin{figure}
\includegraphics[trim={35 25 30 25}, scale=0.64]{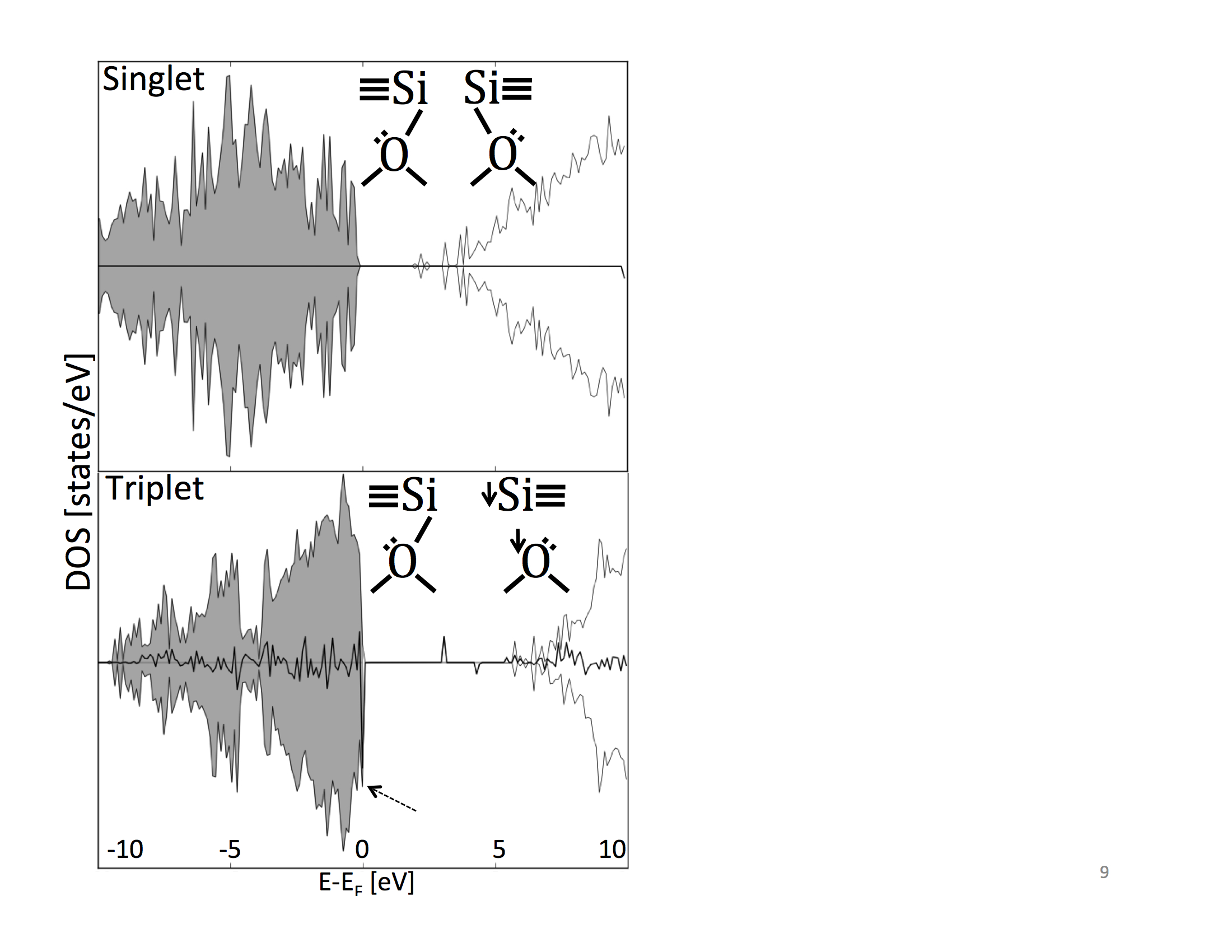}
\caption{\label{nonmag} Spin DOS for the (001) subsurface $V_{\O}^{2+}$ in the (a) singlet and (b) triplet configurations. The triplet state has much higher energy than the singlet state, due to the stability of the electrons in the bond between Si and III-O compared to the dangling bond of the III-Si.  The difference between the spin up versus spin down DOS is shown with the thicker line varying around zero.  
}
\end{figure}

None of the NOV defects have LEMS because the spin up DOS mirrors the spin down DOS and both are equally occupied with a large gap to the excited states. Similarly, when III-O are created from relaxing configurations with III-Si to form new Si--O bonds, with every atom having a complete octet, the spin symmetric state is very stable.   For the III-Si in the (001) subsurface $V_{\O}^{2+}$ and the stable ring defects on the (100) surface, the spin symmetric state is much lower in energy than the magnetic triplet state, as shown representatively in Fig. \ref{nonmag} for the (001) subsurface $V_{\O}^{2+}$ DOS.  In order to access the triplet state, an electron must be removed from the stable Si--O bond and flipped into a previously empty defect state at the top of the valence band [indicated with an arrow in Fig. \ref{nonmag}(b)], which significantly alters the electronic structure and costs around 3~eV. Note that the triplet DOS looks very similar to the DOS in Fig. \ref{DOS} (a) since both have unpaired electrons on an Si and O atom.
 
The Lewis dot structure accompanying the triplet DOS in Fig. \ref{nonmag}(b) is a cartoon of a possible configuration of the two unpaired spin down electrons, derived by visualizing the difference in spin up versus spin down charge density.  The electron at the top of the valence band is localized near a Si atom, but the other electron is delocalized among many oxygen atoms. The singlet and triplet DOS for the ring $V_{\O}^{2+}$ on the (100) surface look very similar to the singlet and triplet DOS given in Fig. \ref{nonmag} because the ring $V_{\O}^{2+}$ also forms two new Si--O bonds, resulting in 2 III-O.   Forming III-O removes charge density from the top of the valence band and increases the charge density in DOS $\sim$4~eV below the valence band maximum (VBM). 
 
 \subsection{\label{sec:level8}Magnetic defects}
 
In contrast to the non-magnetic defects, the low energy defect state from the III-Si atom is easily populated with one electron to give a local magnetic moment, even if the total magnetization is zero. The empty Si--Si bond on the (001) surface and the dangling bond of the Si pointing out of the (100) surface, highlighted with blue isosurfaces in Fig. \ref{fig:empty}, both have low energy defect states that will accept a single electron. All the $V_{\O}^{1+}$ have magnetic ground states, as expected from the odd number of electrons.  

\begin{figure}
\includegraphics[trim={180 10 150 0}, scale=0.62]{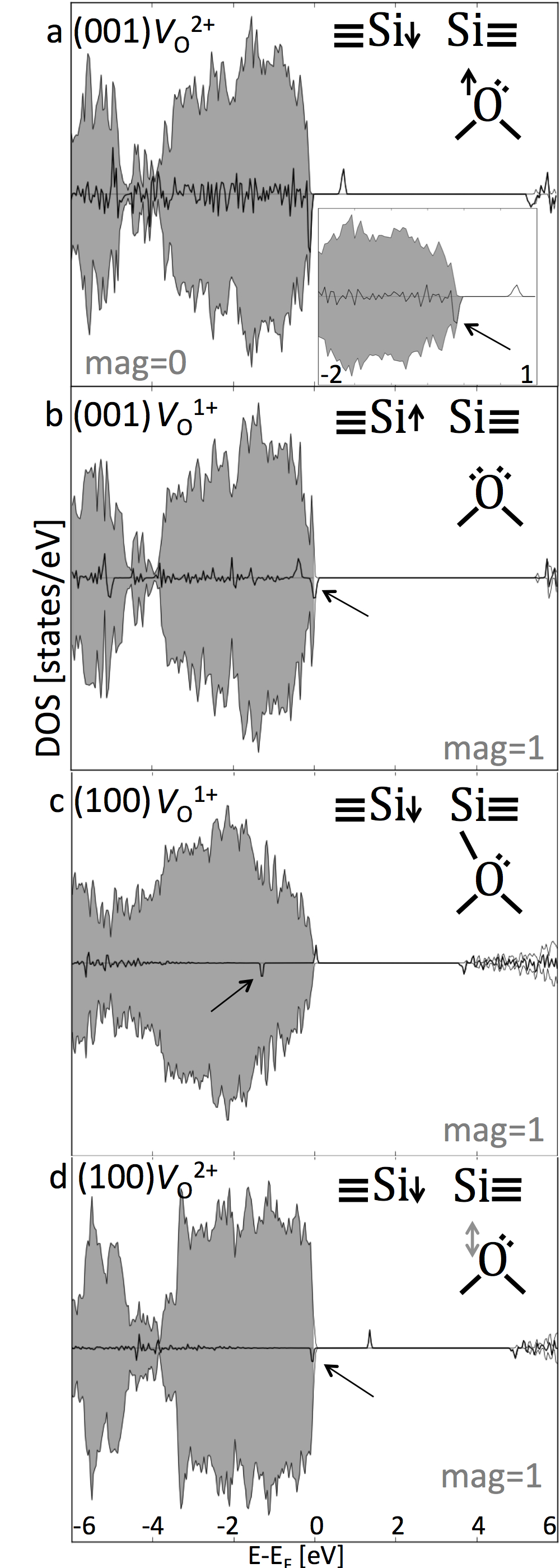}
\caption{\label{DOS} Spin DOS for several additional representative oxygen deficiency defects, similar to Fig. \ref{nonmag}. (a) A close up of the DOS near the top of the valence band for the non-magnetic $V_{\O}^{2+}$ on the (001) surface shows a local magnetic moment.  (b) The $V_{\O}^{1+}$ on the (001) surface shows no bond reformation. (c) The (100) bridge $V_{\O}^{1+}$ forms one extra Si--O bond.  (d) The (100) bridge $V_{\O}^{2+}$ is magnetic. }
\end{figure}

The localized defect state of the singly occupied Si--Si bond on the (001) surface is almost degenerate in energy with the VBM, as shown in Fig. \ref{DOS}(a) for the 2+ charge state and (b) for the 1+ charge state.  The top of the valence band is mainly derived from oxygen lone pairs, so the energy of the electron in the Si--Si bond is similar to the oxygen lone pair electrons.  When two electrons occupy the Si--Si bond, as in the symmetric DOS for the NOV, the energy of the defect states are about 0.5~eV above the valence band.  

In the Lewis dot structure from Fig. \ref{DOS}(a) for the ground state of the $V_{\O}^{2+}$ on the (001) surface, the integral of the spin up minus spin down DOS (total net magnetization) is zero, though this is difficult to see by inspection of the asymmetric DOS. An unpaired spin down electron near the VBM is indicated with the arrow in the inset of Fig. \ref{DOS}(a), which shows a close-up of the (VBM).  In this nominally non-magnetic state, the electron can lower its energy by 0.01~eV by participating in the Si--Si bond instead of residing in the oxygen 2$p$ valence band states, leading to a local magnetic state.   While this DOS is similar to the (001) subsurface triplet DOS in Fig. \ref{nonmag}, the geometries are very different.

We stipulate that, similar to Hund's rule for paramagnetic oxygen molecules, the unpaired electron experiences increased stability in the Si--Si bond due to less nuclear screening than as a paired electron in an oxygen lone pair orbital.  Plotting the charge density difference between the spin up and down densities shows that the extra spin down electron is localized in the Si--Si bond [like in Fig. \ref{fig:empty}(b)], but that the spin up electron is delocalized among many oxygen lone pair orbitals deeper in the slab. This leads to a localized magnetic moment at the surface, which can easily flip spin, such that the energy difference between the 0 versus 2 $\mu_{B}$ magnetization is very small.
  
In order to achieve a magnetization of 2 $\mu_{B}$, the electron must flip to the empty spin up state, which is a little less than 1~eV above the VBM.  However, once the up state is populated, it drops in energy much closer to the top of the valence band and the empty state moves away such that the magnetic excited state is only about 0.001~eV higher in energy than the non-magnetic state, as plotted in Fig. \ref{fig:mag}.  Changes in the defect's magnetic moment, either magnitude or direction,\cite{lee13} can lead to magnetic flux fluctuations during the operation of a superconducting device \cite{choi09}.

We find generally that all the defects with LEMS have asymmetric spin up and down channels, even if the total magnetization of the slab is zero. The simulation cells with an odd number of electrons lend insight to the magnetic properties of cells with even number of electrons,  so we compare the odd electronic structures first.  The non-magnetic excited state is more easily accessed in the Si--Si bond on the (001) surface than in the PC on the (100) surface for the $V_{\O}^{1+}$. The $V_{\O}^{1+}$ on the (001) surface in Fig. \ref{DOS}(b) shows that the spin up and spin down defect states both sit at the top of the valence band, but only the spin up state is occupied.  As mentioned above, the non-magnetic defect structure is around 0.04~eV higher in energy than the magnetic defect structure.  In order to achieve the non-magnetic $V_{\O}^{1+}$ state, only small changes in occupation of states at the top of the valence band are necessary, thus leading to LEMS.  

In contrast, the non-magnetic excited state of the (100) $V_{\O}^{1+}$ is 0.22~eV higher in energy than the magnetic ground state.  As can be seen from Fig. \ref{fig:empty}, the III-Si points out of the surface in the PC, while the other Si atom that used to be bonded to the vacant oxygen has formed a new Si--O bond.  In the 1+ charge state, the extra spin down electron sits deep in the valence band, while an empty spin up state sits near the top of the valence band, as shown in Fig. \ref{DOS}(c). Note also how the entire electronic structure has changed to reflect the new Si--O bond of the III-O [compare the shape of the DOS in Fig. \ref{DOS}(c) to those in Figs. \ref{DOS}(a),(b),(d)].  The non-magnetic $V_{\O}^{1+}$ on the (100) surface is much higher in energy than the magnetic ground state because the spin down state sits much deeper in the valence band for the (100) PC.  

Accessing either of these non-magnetic excited state structures requires much more energy than is available below the 1.2 K superconducting transition temperature of an Al superconducting qubit \cite{cochran58}, so it is difficult to directly connect them to thermal magnetic fluctuations.  However, these defects could still couple to form spin clusters that could contribute to the source of the magnetic noise \cite{sendelbach09}.  

In general, even if there is an even number of electrons in the system, an electron can lower its energy by populating an empty III-Si orbital, giving a local magnetic moment.  Thus, as shown in Fig. \ref{DOS}(d), the ground state of the bridge $V_{\O}^{2+}$ has a total local magnetic moment of 1 $\mu_B$ due to the unpaired electron which is noted with an arrow.   The Lewis dot structure shows the spin down electron localized on the III-Si atom and schematically indicates that the other unpaired electron's charge density is delocalized over the spin up and down channels with the grey double headed arrow.

\subsection{\label{sec:level9}SC qubit noise source}

The III-Si defects have the potential to cause magnetic flux fluctuations if they are found on the surface of silica, and in fact, the III-Si is a common defect that has been identified in both bulk silica and $\alpha$-quartz from first-principle calculations and experiments \cite{andersonPRL11, andersonAPL12, trisi}.  Much effort has gone into determining the reaction pathway for III-Si atoms trapping holes because the neutral III-Si is a precursor to the paramagnetic E$^\prime$-type defect in both amorphous and crystalline.  Both the Si--Si dimer which has lost one electron and the PC dangling bond have been identified as common paramagnetic defects with a trapped hole\cite{andersonPRL11}.  It is worth noting that the nominal charge states of the PC $V_\O$ in our calculations may differ from the typical way the PC defect is described in the literature, as we detail below. 

In our calculations, an empty dangling bond is created by removing two electrons and thus is notated with a 2+ charge, but in first-principles studies of silica and $\alpha$-quartz, the isolated paramagnetic III-Si defect can be created in stoichiometric cells as well.  This defect has one unpaired electron in the dangling bond and is considered to be neutral because the Si has all 4 valence electrons.  Upon hole trapping, the III-Si obtains 1+ charge and an empty dangling bond.  Thus, we consider our $V_{\O}^{2+}$ bridge defect to have a similar electronic structure as a PC III-Si that has trapped a hole. Furthermore, we calculate only a small difference in $E_{\text{form}}$ between the 1+ and 2+ charge states, so trapping a hole to give an empty Si dangling bond does not require significant energy.   Additionally, the E$^\prime$ type structures have been measured in the triplet states\cite{boeroPRL03}, which indicates that perhaps these experiments have already detected our computed LEMS defects with magnetization of 2~$\mu_{\text{B}}$.  

A model SQ system with a density of 5x10$^{17}$~spins/m$^2$, which is 1~spin/200 \AA$^{2}$, has been shown to produce the experimentally measured 1/f noise\cite{sendelbach08}.  Our simulation cells have defect concentrations of 1~LEMS/95~\AA$^2$ and 1~LEMS/150~\AA$^2$ for the (001) and (100) surfaces respectively, which is only slightly higher than the experimentally measured spin density. Given a sparser density of LEMS in our models, we do not expect our results to change because we are already approaching the dilute limit.  We have tested finite size effects using two different (100) supercells, with surface areas of 214 \AA$^2$ versus 95 \AA$^2$ per side.  We find no significant difference in the formation energy and magnetic stability of the NOV defect between these two cells.  Thus, our simulation cells represent a reasonable model of the experimentally predicted spin density of 1 spin/200 \AA$^2$.  

\section{\label{sec:conc}Conclusions}

In summary, we have gained an atomistic understanding of the magnetic instability of oxygen deficiency centers on $\alpha$-quartz surfaces. The empty orbitals associated with III-Si atoms allow electron density transfer between the spin up and down channels, causing low energy magnetic ground or excited states (LEMS).  These under-coordinated Si atoms in dimer and PC configurations are experimentally known to be common in both stoichiometric and oxygen deficient silica and quartz, making them probable candidates for a source of magnetic flux fluctuations in SQs and SQUIDs.  Ideally, our study would reveal a path to remove the source of magnetic noise from silica substrates, and in fact, annealing silica can reduce the density of paramagnetic ODCs \cite{shen13,griscom85}.  
As these defects are intrinsic to silica and cannot be completely eliminated from an amorphous material, reducing magnetic flux fluctuations by engineering the silica surface is unlikely.   Additionally, our computation of low energy magnetic states due to dangling-bonds should be of interest to any researcher attempting to control magnetism at the atomic level, such as spintronics developers\cite{zutic04}.

\begin{acknowledgments}
This work performed under the auspices of the U.S. Department of Energy by Lawrence Livermore National Laboratory under Contract No. DE-AC52-07NA27344
with support from the Laboratory Directed Research and Development Program, tracking number12-ERD-020.
\end{acknowledgments}

\bibliography{mainfile}{}

\end{document}